\documentclass[aps,prl,twocolumn,superscriptaddress,showpacs,showkeys]{revtex4-1}

\usepackage{bm}
\usepackage{graphicx}
\usepackage[usenames]{xcolor}
\usepackage{hyperref}

\begin{document}


\newcommand{\real}{\text{Re}}
\newcommand{\e}{\text{e}}



\title{Dykes for filtering ocean waves using c-shaped vertical cylinders.}


\author{Guillaume Dupont}
\email[Corresponding author: ]{dupont.guillaume@centrale-marseille.fr}
\author{Fabien Remy}
\author{Olivier Kimmoun}
\author{Bernard Molin}
\affiliation{Aix Marseille Univ, CNRS, Centrale Marseille, IRPHE UMR 7342 , 13013 Marseille,
France}

\author{Sebastien Guenneau}
\author{Stefan Enoch}
\affiliation{Aix Marseille Univ, CNRS, Centrale Marseille, Institut Fresnel UMR 7249 , 13013 Marseille,
France}


\date{\today}

\begin{abstract}
The present study investigates a way to design dykes which can filter the wavelengths of ocean surface waves. 
This offers the possibility to achieve a structure that can attenuate waves 
associated with storm swell, without affecting coastline in other conditions. 
Our approach is based on low frequency resonances in metamaterials combined 
with Bragg frequencies for which waves cannot propagate in periodic lattices. 
\end{abstract}


\maketitle



Over the last decades, major natural disasters due to large storms have impacted coastal zones, 
leading to important flooding with material and human losses. 
The water levels associated with storms, at coastal locations, are combination 
of storm surges, local tides and storm waves.
Storm waves (storm swell) which can be characterized by large significant wave heights 
and long wave periods   
take an important contribution in the effect of storms on coastline \cite{noaa}.

Under those observations, we propose a way to consequently reduce the contribution 
of storm waves, in order to protect coastal zones.\\ 

Our study is built upon the works on periodic structures in the area of metamaterials 
and more specifically on low frequencies band-gaps \cite{photoniccrystals}.

In the context of metamaterials for water waves, some works have been done 
in the last years in order to render objects invivisible to waves. 
This cloaking approach has allowed to propose 
some developments to protect structures or areas in open sea 
\cite{farhat:prl2008,alam:prl2012,Newman:EJM2014,PorterCloak:JFM2014,zareei:jfm2015,dupont:jfm2016,Berraquero:pre2013,Dupont:pre2015} 
for most of the cases, but also to control the trajectories of waves. 
Regarding the control of water wave trajectories, periodic structures present some interesting properties.
The study of such structures in solid state physics and optics 
has emerged in the area of photonic crystals \cite{photoniccrystals}, 
and has been translated in water wave physics 
\cite{linton1996scattering,mciver2000water,hu2004pre,huchan2005prl,peter2006water,farhat2010all,McIver:QJMAM2014}.\\

In our study, we propose to use a periodic array of C-shaped cylinders, 
also known as split-ring resonators for artificial magnetism in optics \cite{pendry:srr}. 
Such a lattice has already been used to observe negative effective gravity \cite{huchan2013}.
However, another important feature of this kind of lattice is 
to have a low frequency band-gap 
\cite{movchanguenneau:prb2004}, 
at variance with plain cylinders arrays. 
Thus, it is possible to define a periodic array with constituants of reasonable size with regards 
to wavelength that can attenuate waves associated with storm swell 
(where the period is increased relative to mild sea state) 
but not the waves associated with normal sea state. 
This means that the coastline is not impacted by the presence of the structure in the case of 
normal sea state but that the structure is ``active'' only when the wave period is long.\\
 
In this letter, we demonstrate numerically and experimentally in the case of linear water wave theory 
the efficience of a periodic lattice composed of C-shaped cylinders to protect coastal zone. In addition, 
we show that such a structure remains efficient when the amplitude of waves increases 
(i.e. when the nonlinearities become non negligible).\\


We consider a fluid domain of waterdepth $h$ with an infinite periodic lattice of rigid, 
vertical and bottom mounted objects. 
The Cartesian coordinate system is chosen with $x$ and $y$ as horizontal directions 
and $z$ the vertical upward direction with origin taken at the mean free surface. 
We assume linearised potential flow theory of water waves, 
where the fluid is taken as inviscid, 
incompressible with irrotationnal flow, to describe the propagation of water waves 
with angular frequency $\omega$ through the array. 
Under these assumptions, the velocity potential can be written as
\begin{equation}
\Phi(x,y,z,t)=\real\left\{\phi(x,y)\frac{\cosh\left(k(z+h)\right)}{\cosh(kh)}\e^{-i\omega t}\right\}.
\end{equation}
$\phi$ satisfies the Helmholtz equation
\begin{equation} \label{eq:helm}
 \left(\nabla^{2}+k^{2}\right)\phi=0
\end{equation}
in the fluid domain, with $\nabla=\left( \partial/\partial x\,,\,\partial/\partial y\right)^T$ 
and the wavenumber $k$ solution of the dispersion equation
\begin{equation}
 \omega^2=gk\tanh(kh)
\end{equation}
with $g$ the gravitationnal acceleration.

In addition, Neumann boundary conditions 
\begin{equation}\label{eq:neum}
 \nabla \phi . \bm{n} = 0
\end{equation}
are assumed on the rigid object surfaces, where $\bm{n}$ is outward unit normal vector.\\
\begin{figure}[h]
\centering
\includegraphics[width=0.45\textwidth]{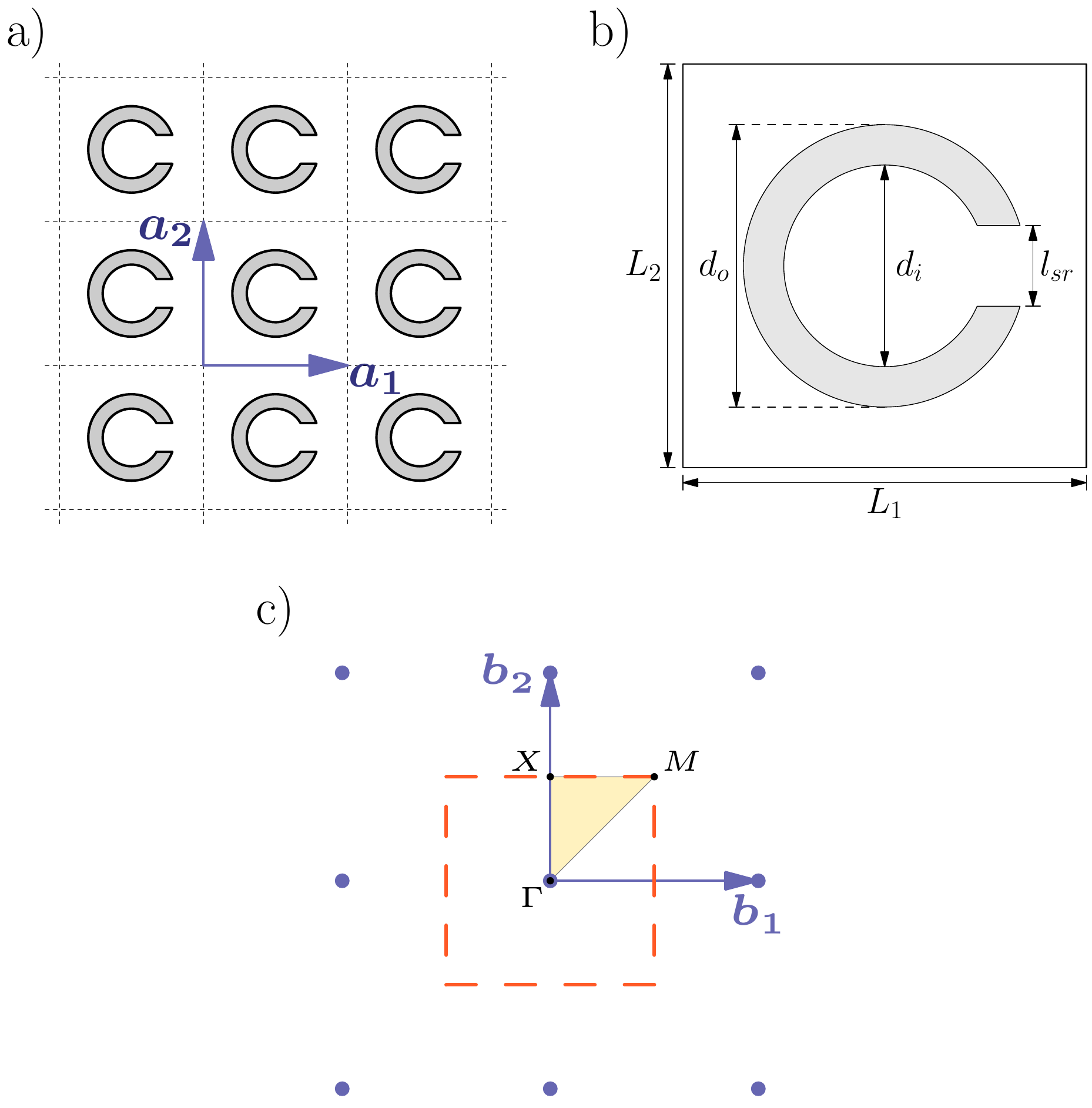}
\caption{\label{fig:latC}
a) Top view of a 2D rectangular lattice with lattice vectors $\bm{a_1}$ and $\bm{a_2}$, 
composed of C-shaped cylinders. 
b) Periodic cell $[L_1 \times L_2]$ of the lattice spanned by $\bm{a_1}$ and $\bm{a_2}$, 
with $d_o$ the outer diameter, $d_i$ the inner diameter and $l_{sr}$ the width of the slit.     
c) Reciprocal lattice spanned by $\bm{b_1}$ and $\bm{b_2}$. 
The dots represent centers of the pattern, the $1^{st}$ Brillouin Zone is delimited by 
the orange dashed contour and the Irreductible Brillouin Zone the vertices $\Gamma,\,X,\,M$ (yellow area).}
\end{figure}
The lattice consists of an array of C-shaped cylinders defined 
by the two lattice vectors $\bm{a_1}$ and $\bm{a_2}$ (see figure \ref{fig:latC}a)
so that it can be represented by all the translations $\bm{R}$ of a periodic pattern 
displayed in figure \ref{fig:latC}b, with
\begin{equation}
 \bm{R}=n_1\,\bm{a_1}+n_2\,\bm{a_2}
\end{equation}
where $n_1$ and $n_2$ are integers.

Following the theory used in the study of crystal structures 
in solid state physics \cite{kittel} and in 
photonic crystals area \cite{photoniccrystals}, 
we look for solutions of the problem (\ref{eq:helm}) 
using Bloch waves decomposition. 
In this context, a solution $\phi$ of (\ref{eq:helm}) 
can be written as the product of plane wave with 
a function $\psi$ which has the periodicity of the lattice: 
\begin{equation}
 \phi(\bm{r})=\e^{i\bm{q}.\bm{r}}\psi(\bm{r}) \quad \text{with} \quad \psi(\bm{r})=\psi(\bm{r}+\bm{R})
\end{equation}
which is equivalent to 
\begin{equation} \label{eq:bloch}
 \phi(\bm{r}+\bm{R})=\e^{i\bm{q}.\bm{R}}\phi(\bm{r})
\end{equation}
where $\bm{q}$ is the Bloch (real-valued) vector defined in the Brillouin Zone (BZ).
The BZ defines a rectangular cell in reciprocal space, 
$BZ=[-\pi/L_1\,\text{;}\,\pi/L_1]\times [-\pi/L_2\,\text{;}\,\pi/L_2]$, 
and can be further reduced to a right-angle triangle $\Gamma X M$ 
called the Irreductible Brillouin Zone (IBZ), with vertices 
$\Gamma\,(0\,\text{;}\,0)$, $X\,(0\,\text{;}\,\pi/L_2)$, $M\,(\pi/L_1\,\text{;}\,\pi/L_2)$ 
(see figure\ref{fig:latC}c).

Solving the eigenvalue-problem (\ref{eq:helm}) with the condition (\ref{eq:bloch}) 
amounts to defining values of $\bm{q}$ on the boundaries IBZ and solving for all $\bm{q}$. 
This method leads to non-trivial eigensolutions $\phi$ for any pair $(k,\bm{q})$ which are plotted 
on a band diagram.\\


In the present study we consider square lattices with $L_1=L_2=L$, 
composed of rigid vertical and bottom mounted cylinders with C-shape (see figure \ref{fig:latC}a-b). 
We study three different configurations 
where geometric parameters are reported in table \ref{tabl:cfgsr}, 
with $H/\lambda$ the waves steepness ($H/\lambda=2a/\lambda$ with $a$ and $\lambda$ 
respectively the amplitude and the wavelength of incoming waves). 
The configuration $2$ is quite close to the configuration $1$ 
where all the parameters have been multiplied by a factor $\sim 2$. 
This will illustrate that the approach is valid from deep to shallow water.
We note that the only difference between configurations $2$ and $3$ 
is the size of the slit $l_{sr}$ .\\

\begin{table}[h]
\centering
\begin{tabular}{c|c|c|c|c|c}
  & $L\,(\text{m})$ & $d_i\,(\text{m})$ & $d_o\,(\text{m})$ & $l_{sr}\,(\text{m})$ & $H/\lambda\,(\%)$\\
  \hline
 cfg 1 & $0.325$ & $0.144$ & $0.150$ & $0.025$ & $2\,\text{;}\,5$  \\
   \hline
 cfg 2 & $0.650$ & $0.290$ & $0.300$ & $0.050$ & $2$ \\
   \hline
 cfg 3 & $0.650$ & $0.290$ & $0.300$ & $0.160$ & $2\,\text{;}\,3.5$
 \end{tabular}  
 \caption{\label{tabl:cfgsr}
 Set of geometric parameters corresponding to the studied configurations.}
\end{table}

\begin{figure}[h]
\centering
 \includegraphics[width=0.45\textwidth]{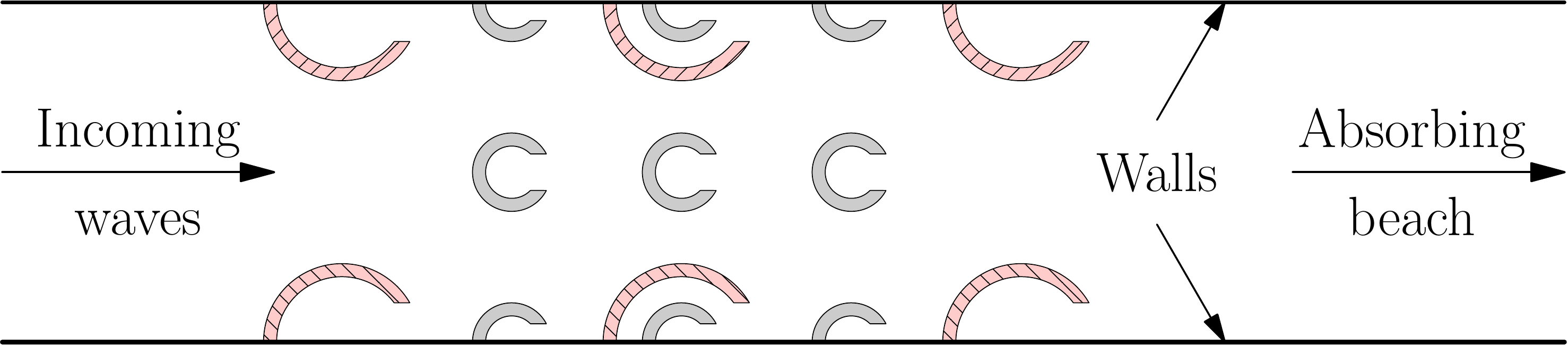}
 \caption{\label{schem:exp}Schematical top view of the experimentally tested arrays. 
 Array with smaller size of C-shape cylinder (gray) corresponding to configuration 1 in table \ref{tabl:cfgsr}. 
 Array with larger size of C-shape cylinder (red with hatch) corresponding to configurations 2 and 3.}
\end{figure}

\begin{figure}[h]
 \centering
\includegraphics[width=0.4\textwidth]{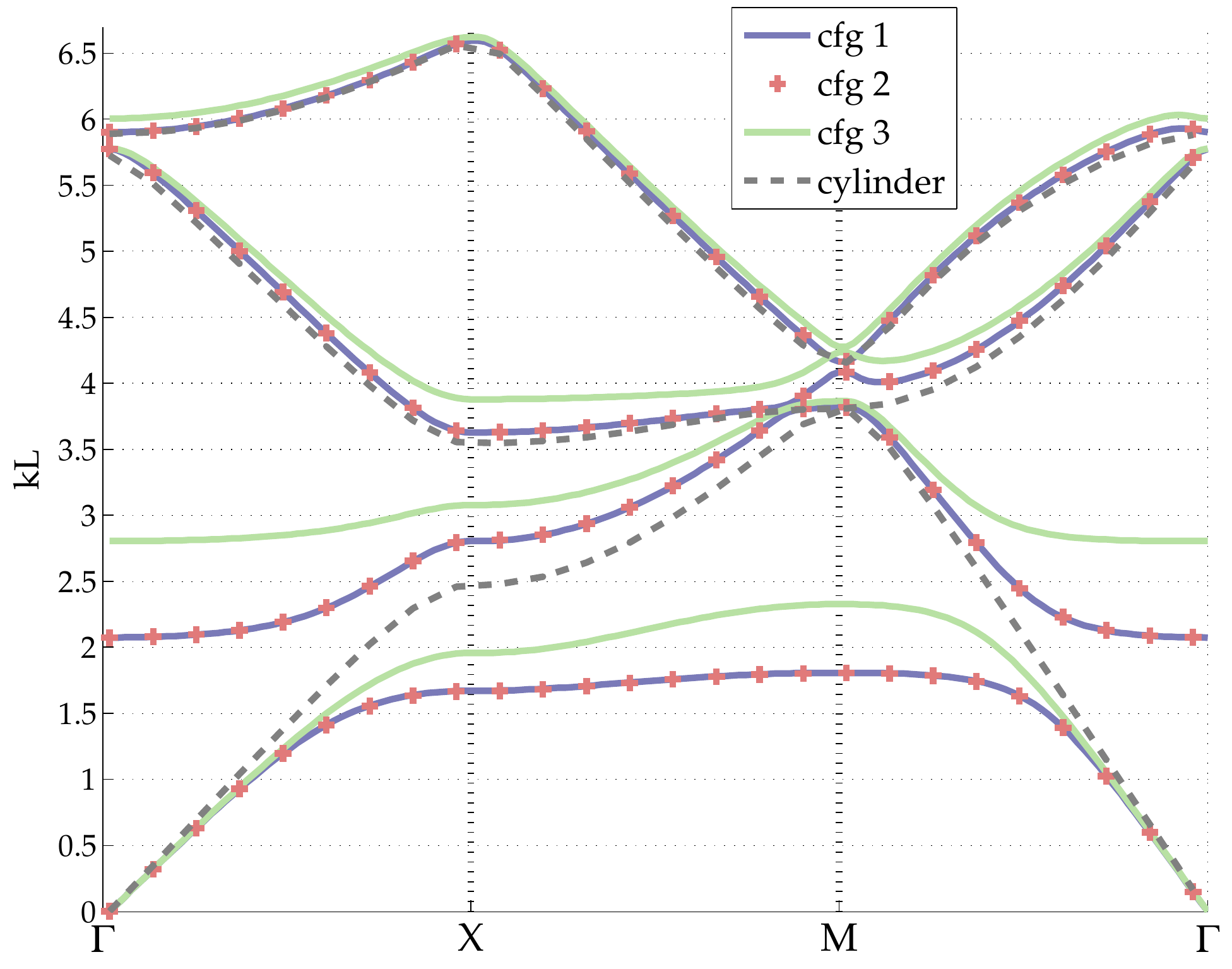} 
\caption{\label{fig:bandsr}
Band diagrams corresponding to the configurations mentionned in table \ref{tabl:cfgsr}. 
The dashed lines stand for bands associated with a square lattice $L=0.325\,\text{m}$ 
of plain cylinders with diameter $d_o=0.150\,\text{m}$.}
\end{figure}

As a first step, the eigenvalue problem given by Eqs.~(\ref{eq:helm},\ref{eq:neum}) 
is solved using finite element (FE) Galerkin's method, for Bloch waves,
for each of the configurations. The results are shown in figure \ref{fig:bandsr}, 
according to the normalized wavenumber $kL$. 
We first notice the quasi-identical results for configurations $1$ and $2$, 
which means that a scale factor on the lattice is passed on the wavenumber, that is on the wavelength.
For all cases, a band-gap is observed between the first two bands corresponding to low frequencies,  
which is due to the presence of the slit. 
This behaviour is not observed in the case of plain cylinders (see figure \ref{fig:bandsr}, dashed curves).
For the inner and outer diameters we choose, we note the larger the slit (configuration 3), 
the larger the first stop band and the thinner the second one.

The $\Gamma-X$ part of the diagram corresponds to waves propagating along principal directions 
($\bm{a_1}$, $\bm{a_2}$) in the array. 
It appears on this part, two significant partial stop-bands respectively between 
the $1^{st}$ and $2^{nd}$ bands and between the $2^{nd}$ and $3^{rd}$ bands. 
It means that 
waves with frequencies in those stop-bands are disallowed to propagate in the array along 
principal directions. Or alternatively, 
waves with frequencies in those stop-bands coming from free-space normally 
to such a periodic array are reflected and do not propagate through the array.
Moreover, for some frequencies in the first stop-band, 
waves do not propagate through the array for all angles of incidence.

The previous arguments suggest that a periodic array with C-shaped cylinders can be used as a dyke 
that is capable to stop waves over a significant range of wavelengths and 
with even more efficiency for waves propagating with small angles around a principal direction. \\

To illustrate the purpose, we numerically investigate the interaction of regular waves 
with the arrays whose parameters are given in table \ref{tabl:cfgsr}. 
We consider a numerical wave tank with constant waterdepth $h=0.5\,\text{m}$ 
where lattices organized as in figure \ref{schem:exp} take place. 
Those choices of organization for our lattices are motivated by our experimental setup.
For the configuration 1, we have $kh \in [1.5\,\text{;}\,9.3]$ which corresponds 
to water waves in intermediate and deep water and we have $kh \in [0.7\,\text{;}\,4.7]$ 
for configurations 2 and 3, which corresponds to water waves in intermediate water, 
closed to shallow water for smaller $kh$.
Incoming waves are at normal incidence, 
which corresponds to waves propagate in the $\Gamma-X$ direction.

We use the Helmholtz equation (Eq.~(\ref{eq:helm})) combined 
with a FE Galerkin's method to calculate the free-surface elevation.

\begin{figure}[h]
\centering
 \includegraphics[width=0.45\textwidth]{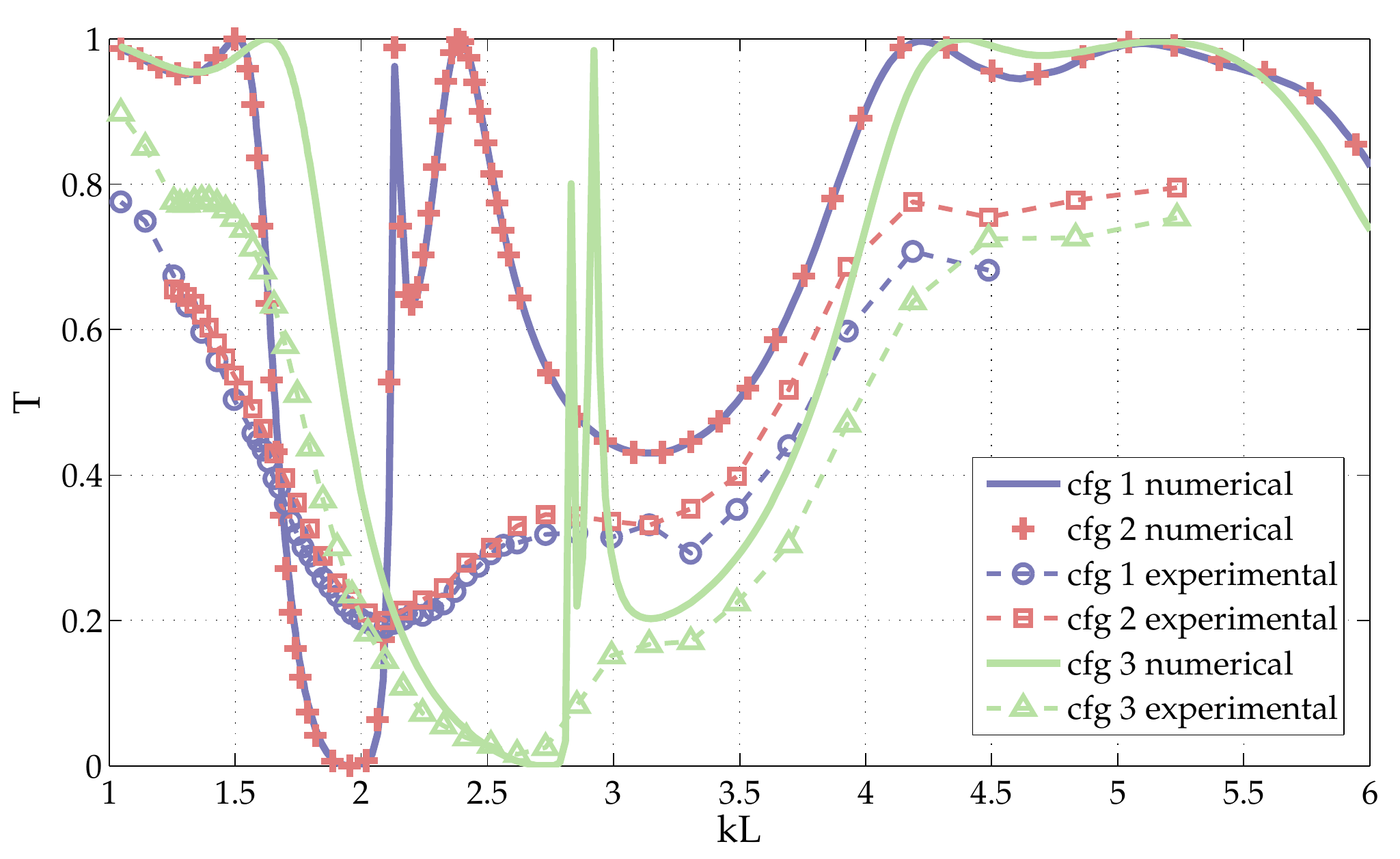}
 \caption{\label{fig:exp1}Comparison of experimental measurements of the transmision coefficient 
 with numerical results for incident water waves with $H\lambda=2\%$. 
 }
\end{figure}

Figure \ref{fig:exp1} shows the transmission coefficients with respect 
to the normalized wavenumber $kL$ for our three configurations. 
As predicted by the band diagrams, results are identical for configurations $1$ and $2$. 
We easily identify the band-gaps at small $kL$ (low frequencies) 
for which the transmission coefficient is null, 
and the stop-bands at larger $kL$ for which the transmission coefficient is low. 
Those two gaps are separated by a peak of transmission corresponding to the second band.\\

\begin{figure}[h]
\centering
 \includegraphics[width=0.45\textwidth]{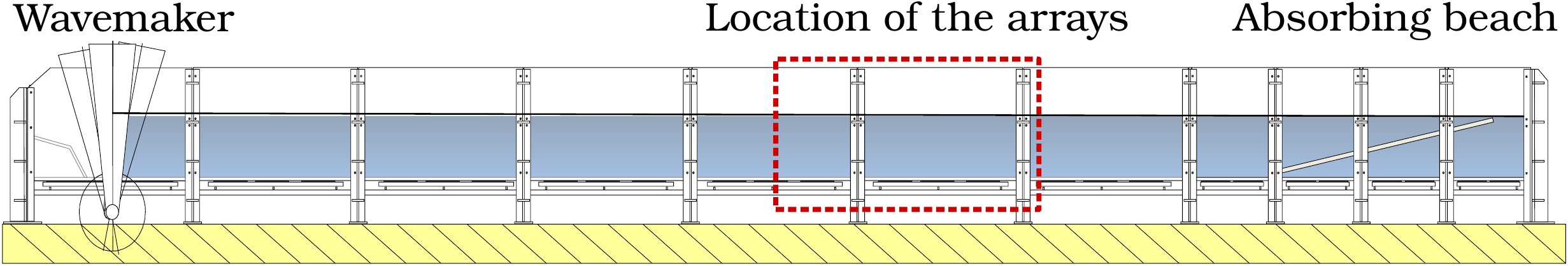}
 \caption{\label{schem:canal}Schematical side-view of the wave tank located 
 at engineering school Centrale Marseille.}
\end{figure}

For the experimental part of the study, we use the wave
tank at engineering school Centrale Marseille, which is $17\,\text{m}$ long and
$0.65\,\text{m}$ wide. Waves are generated by a flap wavemaker, the center 
of the arrays is located at $10\,\text{m}$ from the wavemaker, 
and an efficient $3\,\text{m}$ long absorbing beach 
takes place at the end of the tank (see figure \ref{schem:canal}).

The amplitudes of reflected and transmitted waves are measured with sets of resistive wave gauges 
placed in front and behind the arrays. Additionally, we put a wave gauge inside a cylinder of each column 
of the array.
Each configuration is tested for different wave steepness 
($H/\lambda$ entries in table \ref{tabl:cfgsr}). 

Figure \ref{fig:exp1} shows the measured transmission coefficients for incident 
wave steepness $H/\lambda=2\%$, compared with numerical results. 
An important feature is that we did not observe experimentally the peak of transmission 
corresponding to the second band, for all the configurations.
On the other hand, we note similar results for configurations 1 and 2, 
with a reasonable agreement with  numerical computations, 
in the sense that the global variation is respected. 
Differences with numerics are presumably due to dissipative phenomena, 
such as flow separation induced by sharp corners, viscous effects and wave breaking.
Results for configuration 3, where the size of the slit has been increased, are meanwhile 
in good agreement with numerical results.

\begin{figure}[h]
\centering
 \includegraphics[width=0.45\textwidth]{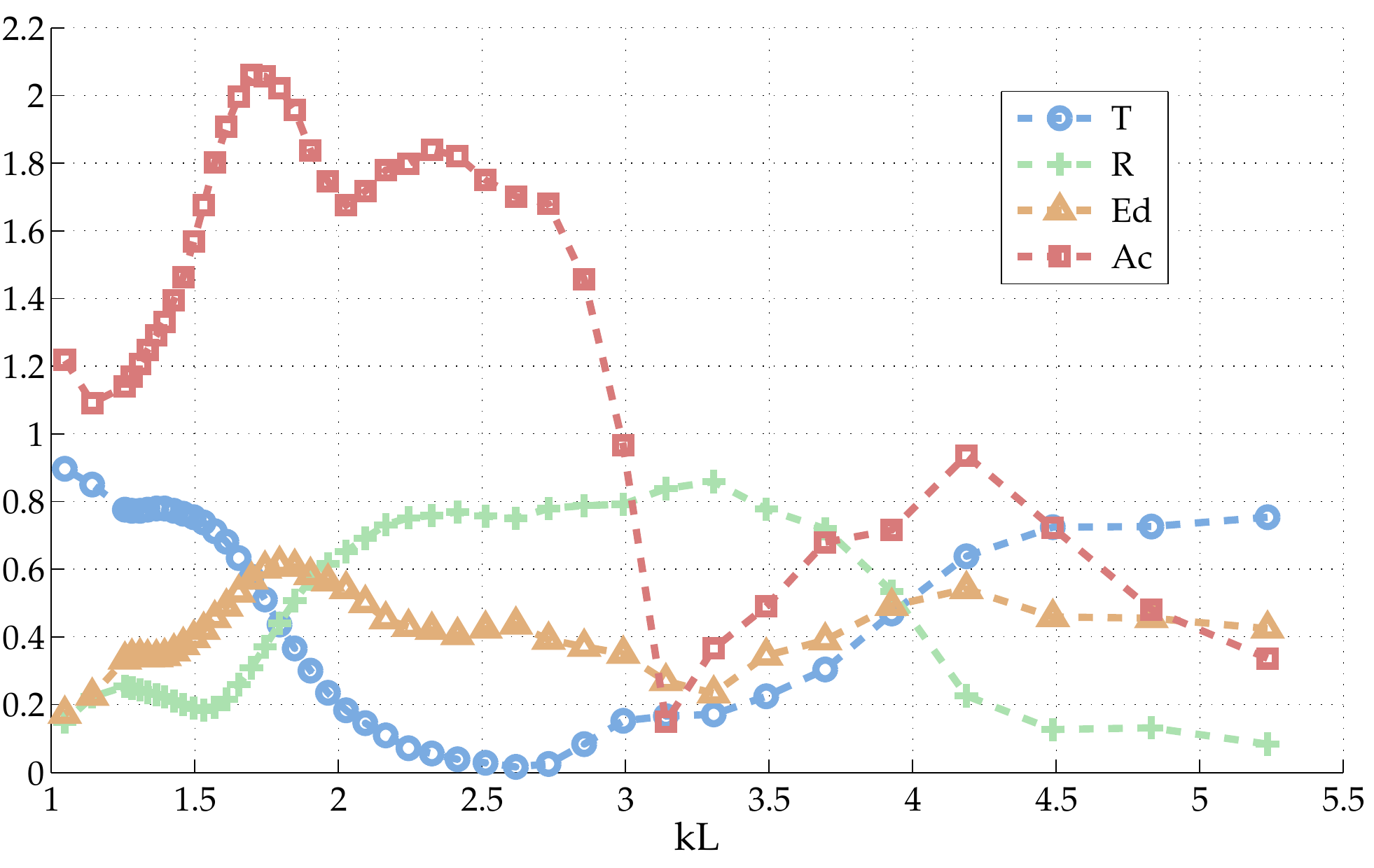}
 \caption{\label{fig:exp2}Experimental measurements of the transmission coefficient (T), 
 reflexion coefficient (R), normalized amplitude inside a cylinder of the $1^{st}$ column (Ac) 
 and dissipation coefficient (Ed), (see Eq.~(\ref{eq:Ed}))
 for configuration $3$.
The amplitude inside the cylinders for other columns is similar but attenuated.}
\end{figure}

\begin{figure}[h]
\centering
 \includegraphics[width=0.45\textwidth]{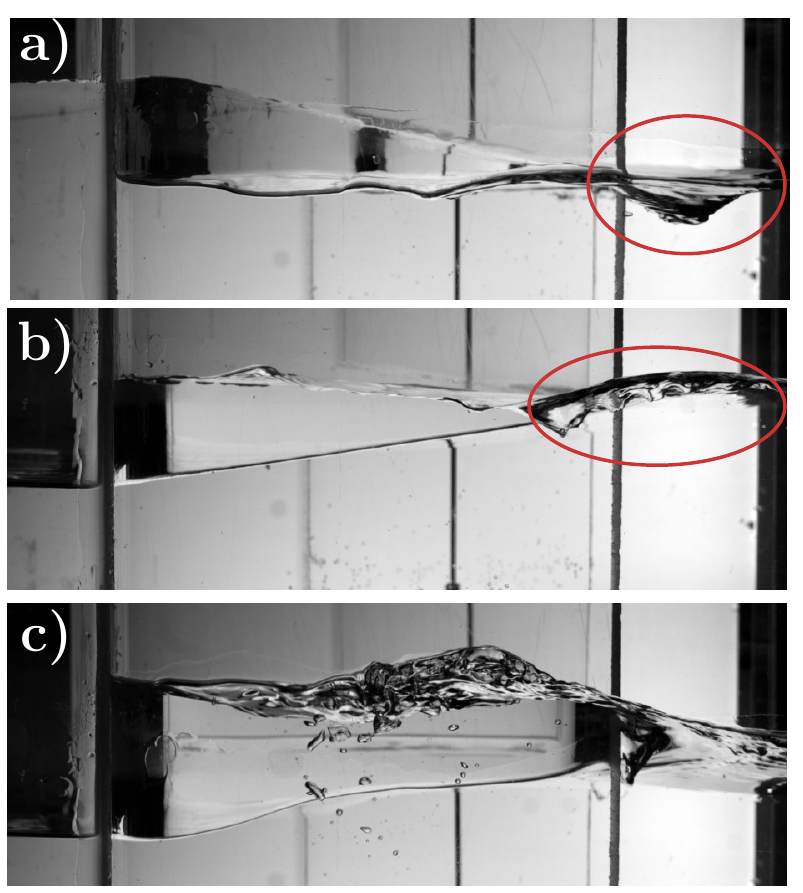}
 \caption{\label{fig:photo1}Photos of a cylinder at different time steps and for 
 different wave steepnesses. a) water comes out of the cylinder for $H/\lambda=2\%$, 
 b) water comes in the cylinder for $H/\lambda=2\%$, 
 c) wave breaking in the cylinder for $H/\lambda=5\%$.
  The vortices are identified inside the red circles.}
\end{figure}

A part of explanation for transmission associated with the second band 
can be found in the measurment of amplitude inside cylinders. 
The results for configuration 3 displayed in figure \ref{fig:exp2} show  
that the normalized amplitude becomes important inside the cylinder for $kL<3$, 
wich includes the second band. 
Additionally, for those values of $kL$, an important flow separation, 
with generation of vortices, 
is observed on both sides of the slit when water waves come in and out the cylinder 
(see figure \ref{fig:photo1}a-b). For values of $kL$ where the amplitude is maximal inside 
the cylinder, wave breaking occurs (see figure \ref{fig:photo1}c).

\begin{figure}[h]
\centering
 \includegraphics[width=0.45\textwidth]{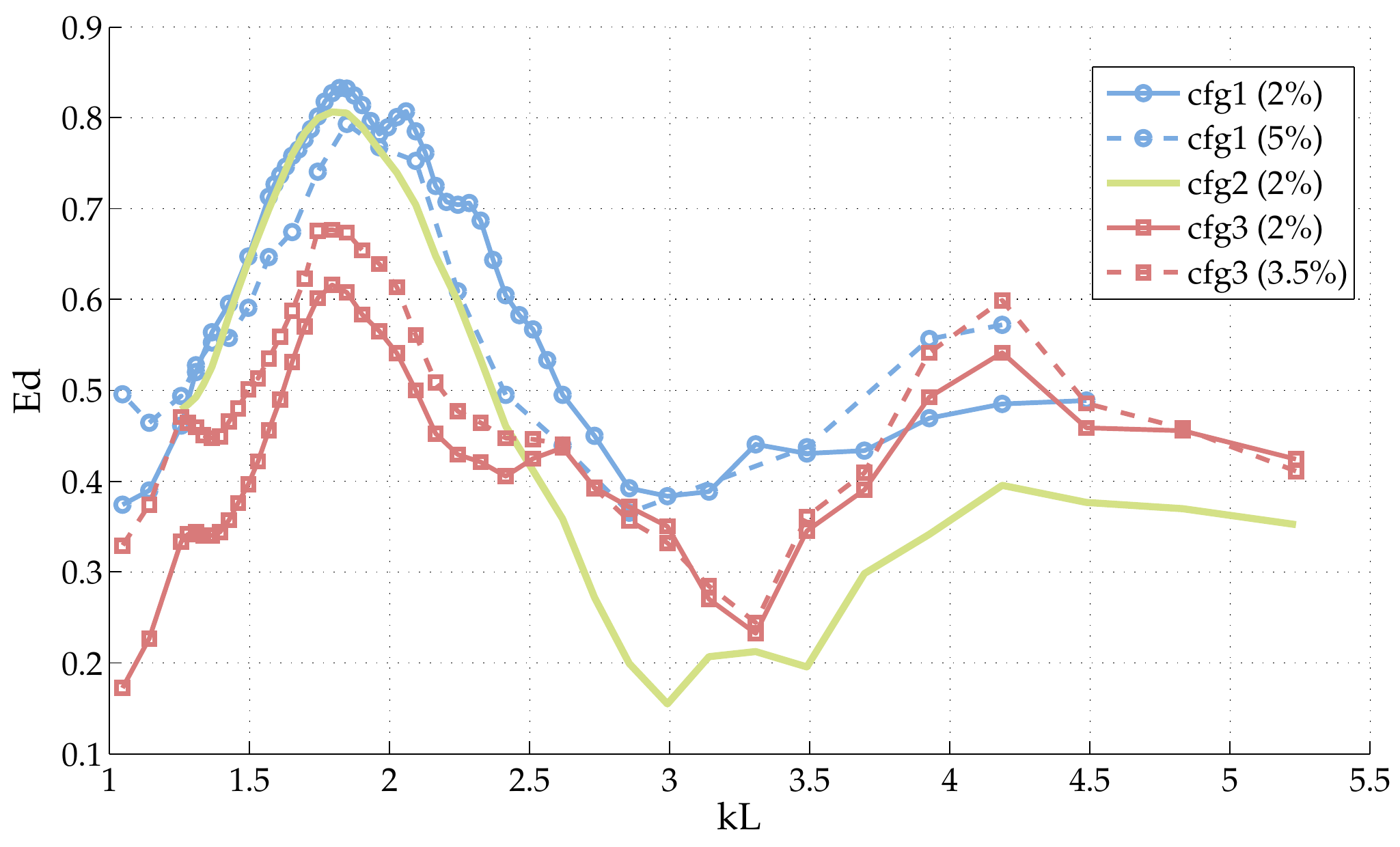}
 \caption{\label{fig:Ed}Dissipation coefficients for all the studied configurations. 
 When the amplitude of waves is important in the cylinder ($kL\lesssim 2.7$), 
 we note that, the thiner the slit (configurations 1,2), the higher the dissipation.}
\end{figure}

The dissipation coefficient 
\begin{equation}\label{eq:Ed}
 E_d=1-\left( R^2+T^2\right)
\end{equation}

which is zero according to linearized potential flow theory,
confirms that when the amplitude becomes significant inside cylinders, dissipation increases 
(yellow curve with triangle markers on figure \ref{fig:exp2}). This fact is observed for all 
the configuration (see figure \ref{fig:Ed}).
Additionnaly, a part of the wave energy is transfered to higher harmonic components. 

\begin{figure}[h]
\centering
 \includegraphics[width=0.45\textwidth]{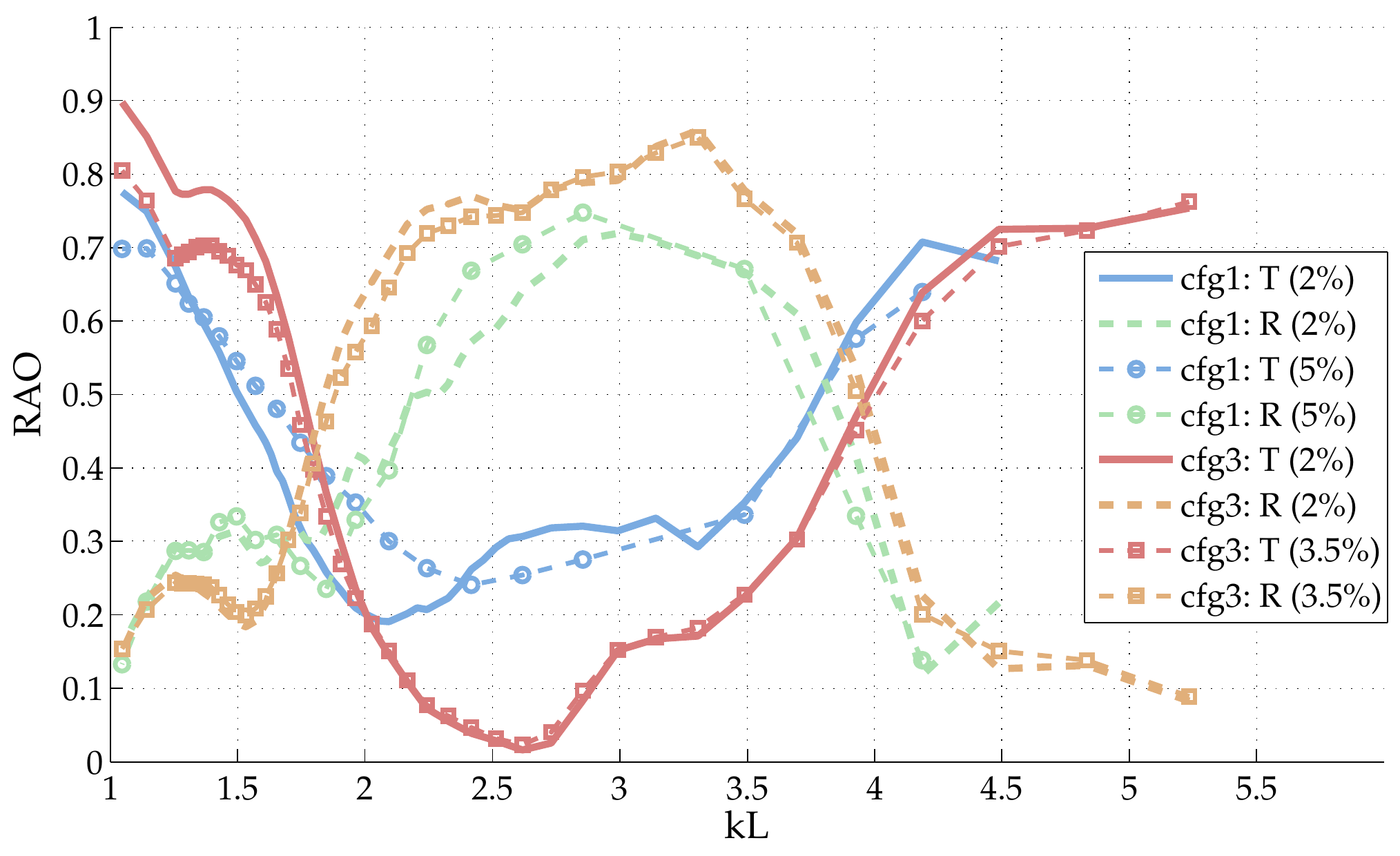}
 \caption{\label{fig:exp3}Comparison of experimental reflexion and transmision coefficients 
 for different waves steepnesses (shown in brackets).
 Results for higher waves steepnesses are represented with markers 
 (circles for configuration 1 and squares for configuration 3).}
\end{figure}

Increasing the amplitude of incident water waves (wave steepness up to $5\%$), 
we obtain results very similar to waves with small amplitudes, as despicted in figure \ref{fig:exp3}. 
Amplitude of transmitted waves can be attenuated by $50\%$ for a wide range of 
wavelengths: $1.5 < kL < 4$ which corresponds to $4L > \lambda > 1.5L$, 
by adjusting the size of the slit. 
More preferably, the attenuation of transmitted waves reaches $70\%$ to $80\%$ for 
$2 < kL < 3.5$ ($3.2L > \lambda > 1.8L$).\\
Those observations evidence that this kind of array makes a good choice
to consequently 
minimize the impact of storm swell on coastal zone.

In this letter, we have performed numerical and experimental demonstration about a 
periodic structure that can be used as a dyke to protect coastal zone.

In a first step, 
we prove numerically 
that our choice of C-shaped cylinders may induce a large low frequency stop band associated with 
a low transmission on a wide range of wavelength. 
In a second step, we illustrate it experimentally and we extend the study to waves 
with large wave heights, 
which confirms the action of the structure on storm swell.

We stress that, with the knowledge of local parameters of a coastal zone 
(bathymetry, periods of waves), it is always possible to define a structure in our way, thanks to the 
link between wavelengths and the lattice parameter $L$. 
Then, for an array defined for waves with $kL\sim 5$ 
(normal sea state, maximum of transmission, invisible structure to water waves), 
a significant increase of the period of waves will be associated with a low transmission 
and a consequent attenuation of the amplitude of waves. 
That means the structure becomes active.
On the other hand, when the period of waves decreases, 
the associated band structure flattens and 
we should not capture thin peaks of transmission. 

Consequently, we are convinced that our results pave the way 
for a new technology of dykes, active for most cases of storm swell. 
This technology may help to preserve environment and is less invasive than classical dykes.

\end{document}